
\magnification=1200

\tolerance=10000



\parskip=\medskipamount
\overfullrule=0pt
\raggedbottom
\def\normalparindent{24pt}
\nopagenumbers
\footline={\ifnum\pageno=1{\hfil}\else{\hfil\rm\folio\hfil}\fi}
\def\endpage{\vfill\eject}
\def\beginlinemode{\endmode\begingroup\parskip=0pt
                   \obeylines\def\\{\par}\def\endmode{\par\endgroup}}
\def\beginparmode{\endmode\begingroup \def\endmode{\par\endgroup}}
\let\endmode=\par
\def\raggedcenter{
                  \leftskip=2em plus 6em \rightskip=\leftskip
                  \parindent=0pt \parfillskip=0pt \spaceskip=.3333em
                  \xspaceskip=.5em\pretolerance=9999 \tolerance=9999
                  \hyphenpenalty=9999 \exhyphenpenalty=9999 }
\def\\{\cr}
\let\rawfootnote=\footnote\def\footnote#1#2{{\parindent=0pt\parskip=0pt
        \rawfootnote{#1}{#2\hfill\vrule height 0pt depth 6pt width 0pt}}}
\def\title{\null\vskip 3pt plus 0.2fill\beginlinemode\raggedcenter\bf}
\def\author{\vskip 3pt plus 0.2fill \beginlinemode\raggedcenter}
\def\affil{\vskip 3pt plus 0.1fill\beginlinemode\raggedcenter\it}
\def\abstract{\vskip 3pt plus 0.3fill \beginparmode{\noindent  ABSTRACT:~}  }
\def\endtitlepage{\endpage\body}
\def\body{\beginparmode\parindent=\normalparindent}
\def\head#1{\par\goodbreak{\immediate\write16{#1}
           {\noindent\bf #1}\par}\nobreak\nobreak}

\def\refto#1{$^{[#1]}$}
\def\ref#1{Ref.~#1}
\def\Ref#1{Ref.~#1}\def\cite#1{{#1}}\def\[#1]{[\cite{#1}]}

\def\(#1){(\call{#1})}
\def\call#1{{#1}}\def\taghead#1{{#1}}
\def\references{\head{REFERENCES}\beginparmode\frenchspacing\parskip=0pt}
\gdef\refis#1{\item{#1.\ }}
\gdef\journal#1,#2,#3,#4.{{\it #1~}{\bf #2}, #3 (#4)}
\def\endreferences{\body}
\def\endit{\endmode\vfill\supereject}\let\endpaper=\endit

\def\apj{\journal Ap. J.,}
\def\apjs{\journal Ap. J. Suppl.,}

\def\pasp{\journal P.A.S.P.,}
\def\mnras{\journal M.N.R.A.S.,}

\def\nat{\journal Nature,}
\def\araa{\journal Ann.Rev.Astron.Astrophys.,}

\def\gsim{\mathrel{\raise.3ex\hbox{$>$\kern-.75em\lower1ex\hbox{$\sim$}}}}
\def\lsim{\mathrel{\raise.3ex\hbox{$<$\kern-.75em\lower1ex\hbox{$\sim$}}}}
\def\square{\kern1pt\vbox{\hrule height 0.6pt\hbox{\vrule width 0.6pt\hskip 3pt
   \vbox{\vskip 6pt}\hskip 3pt\vrule width 0.6pt}\hrule height 0.6pt}\kern1pt}
\def\sla{\raise.15ex\hbox{$/$}\kern-.72em}

\catcode`@=11
\newcount\r@fcount \r@fcount=0\newcount\r@fcurr
\immediate\newwrite\reffile\newif\ifr@ffile\r@ffilefalse
\def\w@rnwrite#1{\ifr@ffile\immediate\write\reffile{#1}\fi\message{#1}}
\def\writer@f#1>>{}
\def\referencefile{\r@ffiletrue\immediate\openout\reffile=\jobname.ref%
  \def\writer@f##1>>{\ifr@ffile\immediate\write\reffile%
    {\noexpand\refis{##1} = \csname r@fnum##1\endcsname = %
     \expandafter\expandafter\expandafter\strip@t\expandafter%
     \meaning\csname r@ftext\csname r@fnum##1\endcsname\endcsname}\fi}%
  \def\strip@t##1>>{}}

\def\citeall#1{\xdef#1##1{#1{\noexpand\cite{##1}}}}
\def\cite#1{\each@rg\citer@nge{#1}}
\def\each@rg#1#2{{\let\thecsname=#1\expandafter\first@rg#2,\end,}}
\def\first@rg#1,{\thecsname{#1}\apply@rg}
\def\apply@rg#1,{\ifx\end#1\let\next=\relax%
\else,\thecsname{#1}\let\next=\apply@rg\fi\next}%
\def\citer@nge#1{\citedor@nge#1-\end-}
\def\citer@ngeat#1\end-{#1}
\def\citedor@nge#1-#2-{\ifx\end#2\r@featspace#1
  \else\citel@@p{#1}{#2}\citer@ngeat\fi}
\def\citel@@p#1#2{\ifnum#1>#2{\errmessage{Reference range #1-#2\space is bad.}
    \errhelp{If you cite a series of references by the notation M-N, then M and
    N must be integers, and N must be greater than or equal to M.}}\else%
{\count0=#1\count1=#2\advance\count1 by1\relax\expandafter\r@fcite\the\count0,%
  \loop\advance\count0 by1\relax
    \ifnum\count0<\count1,\expandafter\r@fcite\the\count0,%
  \repeat}\fi}
\def\r@featspace#1#2 {\r@fcite#1#2,}    \def\r@fcite#1,{\ifuncit@d{#1}
    \expandafter\gdef\csname r@ftext\number\r@fcount\endcsname%
    {\message{Reference #1 to be supplied.}\writer@f#1>>#1 to be supplied.\par
     }\fi\csname r@fnum#1\endcsname}
\def\ifuncit@d#1{\expandafter\ifx\csname r@fnum#1\endcsname\relax%
\global\advance\r@fcount by1%
\expandafter\xdef\csname r@fnum#1\endcsname{\number\r@fcount}}
\let\r@fis=\refis   \def\refis#1#2#3\par{\ifuncit@d{#1}%
    \w@rnwrite{Reference #1=\number\r@fcount\space is not cited up to now.}\fi%
  \expandafter\gdef\csname r@ftext\csname r@fnum#1\endcsname\endcsname%
  {\writer@f#1>>#2#3\par}}
\def\r@ferr{\endreferences\errmessage{I was expecting to see
\noexpand\endreferences before now;  I have inserted it here.}}
\let\r@ferences=\references
\def\references{\r@ferences\def\endmode{\r@ferr\par\endgroup}}
\let\endr@ferences=\endreferences
\def\endreferences{\r@fcurr=0{\loop\ifnum\r@fcurr<\r@fcount
    \advance\r@fcurr by 1\relax\expandafter\r@fis\expandafter{\number\r@fcurr}%
    \csname r@ftext\number\r@fcurr\endcsname%
  \repeat}\gdef\r@ferr{}\endr@ferences}
\let\r@fend=\endpaper\gdef\endpaper{\ifr@ffile
\immediate\write16{Cross References written on []\jobname.REF.}\fi\r@fend}
\catcode`@=12
\citeall\refto\citeall\ref\citeall\Ref
\catcode`@=11
\newcount\tagnumber\tagnumber=0
\immediate\newwrite\eqnfile\newif\if@qnfile\@qnfilefalse
\def\write@qn#1{}\def\writenew@qn#1{}
\def\w@rnwrite#1{\write@qn{#1}\message{#1}}
\def\@rrwrite#1{\write@qn{#1}\errmessage{#1}}
\def\taghead#1{\gdef\t@ghead{#1}\global\tagnumber=0}
\def\t@ghead{}\expandafter\def\csname @qnnum-3\endcsname
  {{\t@ghead\advance\tagnumber by -3\relax\number\tagnumber}}
\expandafter\def\csname @qnnum-2\endcsname
  {{\t@ghead\advance\tagnumber by -2\relax\number\tagnumber}}
\expandafter\def\csname @qnnum-1\endcsname
  {{\t@ghead\advance\tagnumber by -1\relax\number\tagnumber}}
\expandafter\def\csname @qnnum0\endcsname
  {\t@ghead\number\tagnumber}
\expandafter\def\csname @qnnum+1\endcsname
  {{\t@ghead\advance\tagnumber by 1\relax\number\tagnumber}}
\expandafter\def\csname @qnnum+2\endcsname
  {{\t@ghead\advance\tagnumber by 2\relax\number\tagnumber}}
\expandafter\def\csname @qnnum+3\endcsname
  {{\t@ghead\advance\tagnumber by 3\relax\number\tagnumber}}
\def\equationfile{\@qnfiletrue\immediate\openout\eqnfile=\jobname.eqn%
  \def\write@qn##1{\if@qnfile\immediate\write\eqnfile{##1}\fi}
  \def\writenew@qn##1{\if@qnfile\immediate\write\eqnfile
    {\noexpand\tag{##1} = (\t@ghead\number\tagnumber)}\fi}}
\def\callall#1{\xdef#1##1{#1{\noexpand\call{##1}}}}
\def\call#1{\each@rg\callr@nge{#1}}
\def\each@rg#1#2{{\let\thecsname=#1\expandafter\first@rg#2,\end,}}
\def\first@rg#1,{\thecsname{#1}\apply@rg}
\def\apply@rg#1,{\ifx\end#1\let\next=\relax%
\else,\thecsname{#1}\let\next=\apply@rg\fi\next}
\def\callr@nge#1{\calldor@nge#1-\end-}\def\callr@ngeat#1\end-{#1}
\def\calldor@nge#1-#2-{\ifx\end#2\@qneatspace#1 %
  \else\calll@@p{#1}{#2}\callr@ngeat\fi}
\def\calll@@p#1#2{\ifnum#1>#2{\@rrwrite{Equation range #1-#2\space is bad.}
\errhelp{If you call a series of equations by the notation M-N, then M and
N must be integers, and N must be greater than or equal to M.}}\else%
{\count0=#1\count1=#2\advance\count1 by1\relax\expandafter\@qncall\the\count0,%
  \loop\advance\count0 by1\relax%
    \ifnum\count0<\count1,\expandafter\@qncall\the\count0,  \repeat}\fi}
\def\@qneatspace#1#2 {\@qncall#1#2,}
\def\@qncall#1,{\ifunc@lled{#1}{\def\next{#1}\ifx\next\empty\else
  \w@rnwrite{Equation number \noexpand\(>>#1<<) has not been defined yet.}
  >>#1<<\fi}\else\csname @qnnum#1\endcsname\fi}
\let\eqnono=\eqno\def\eqno(#1){\tag#1}\def\tag#1$${\eqnono(\displayt@g#1 )$$}
\def\aligntag#1\endaligntag  $${\gdef\tag##1\\{&(##1 )\cr}\eqalignno{#1\\}$$
  \gdef\tag##1$${\eqnono(\displayt@g##1 )$$}}
\def\eqalignno#1{\displ@y \tabskip\centering
  \halign to\displaywidth{\hfil$\displaystyle{##}$\tabskip\z@skip
    &$\displaystyle{{}##}$\hfil\tabskip\centering
    &\llap{$\displayt@gpar##$}\tabskip\z@skip\crcr
    #1\crcr}}
\def\displayt@gpar(#1){(\displayt@g#1 )}
\def\displayt@g#1 {\rm\ifunc@lled{#1}\global\advance\tagnumber by1
        {\def\next{#1}\ifx\next\empty\else\expandafter
        \xdef\csname @qnnum#1\endcsname{\t@ghead\number\tagnumber}\fi}%
  \writenew@qn{#1}\t@ghead\number\tagnumber\else
        {\edef\next{\t@ghead\number\tagnumber}%
        \expandafter\ifx\csname @qnnum#1\endcsname\next\else
        \w@rnwrite{Equation \noexpand\tag{#1} is a duplicate number.}\fi}%
  \csname @qnnum#1\endcsname\fi}
\def\ifunc@lled#1{\expandafter\ifx\csname @qnnum#1\endcsname\relax}
\let\@qnend=\end\gdef\end{\if@qnfile
\immediate\write16{Equation numbers written on []\jobname.EQN.}\fi\@qnend}
\catcode`@=12

\vskip -0.3truein
\rightline{FERMILAB-Pub-93/391-A}
\rightline{December 1993}

\title{Can MACHOs Probe the Shape of the Galaxy Halo?}
\vskip 2pc

\author{Joshua Frieman${}^{1,2}$ \& Rom\'an Scoccimarro${}^3$}
\vskip 2pc

\affil{${}^1$NASA/Fermilab Astrophysics Center \\
       Fermi National Accelerator Laboratory \\
       P. O. Box 500, Batavia, IL 60510}

\affil{${}^2$Department of Astronomy and Astrophysics \\
         University of Chicago, Chicago, IL 60637}

\affil{${}^3$Department of Physics \\
           University of Chicago, Chicago, IL 60637}

\vskip 2pc

\abstract{We study the prospects for the current microlensing
searches, which have recently discovered several candidates,
to yield useful information about the flattening of the
Galaxy dark matter halo. Models of HI warps and N-body simulations
of galaxy formation suggest that disks commonly form in oblate halos
with a tilt between the disk and halo symmetry axes. The microlensing
optical depth for the Large Magellanic Cloud depends sensitively on
the disk-halo tilt angle in the Milky Way.
We conclude that a much larger spread in values for $\tau(LMC)$
is consistent with rotation curve constraints than previously thought,
and that the ratio $\tau(SMC)/\tau(LMC)$ of
the optical depth to the Small and Large Magellanic clouds
is not a clean test of halo flattening.

Subject Headings: dark matter - gravitational lensing - Magellanic Clouds}

\vskip 1pc
Submitted to {\it The Astrophysical Journal Letters}

\endtitlepage

\head{1. Introduction}
\vskip 1pc

The nature of the dark matter in galaxy halos is still unknown, the
current choice being between WIMPs (non-baryonic particles such as axions or
supersymmetric neutralinos) or MACHOs
(massive astrophysical compact halo objects such as brown
dwarfs or more massive stellar remnants), or some combination of the two.
Recently three candidate microlensing events, the presumed signature of
MACHOs in our galaxy, have been observed in the direction of the
Large Magellanic Cloud (LMC) by the MACHO (Alcock et al. 1993)
and EROS (Aubourg et al. 1993) collaborations, and a fourth event
toward the Galactic bulge has been found by the OGLE collaboration
(Udalski et al. 1993). The idea of detecting compact halo objects by
observing the amplification of
stellar magnitude, i.e., gravitational microlensing, when a MACHO
passes near the line of sight to a monitored star was first suggested by
Paczy\'nski (1986).

If MACHOs constitute the bulk of the
halo (see Gates \& Turner 1993), it is clearly of interest to ask
what can be learned about the Milky Way halo from the microlensing
experiments over the next several years. This involves study of how
such quantitites as the microlensing optical depth $\tau$, event rate $\Gamma$,
average event time duration $\langle t_e \rangle$,
and the distributions in event time and lens mass depend on halo
parameters such as core radius and velocity dispersion.
Exploratory studies along these lines were made by
Griest (1991) for spherical halo models and by
Sackett \& Gould (1994, hereafter SG) for flattened halo models.
Imposing constraints from the Galactic rotation curve
and varying assumptions about the Galactic disk and bulge within reasonable
limits, SG found that
the microlensing optical depth for the LMC is relatively insensitive
to the halo model, only varying over the range
$\tau(LMC) \simeq 2 - 5 \times 10^{-7}$. Moreover, they found that
$\tau(LMC)$ is essentially independent of halo flattening while $\tau(SMC)$
is not, and proposed the ratio of optical depths to the small and large clouds,
$\tau(SMC)/\tau(LMC)$, as a robust measure of halo ellipticity, independent
of other unknown halo parameters.

In this Letter, we consider microlensing in a flattened halo from a new
angle, so to speak. Although the shape of galaxy halos is still an open
question (Ashman 1992), several arguments suggest that non-spherical halos may
be the norm for spiral galaxies.
Flattened halos have been invoked
in studies of polar-ring galaxies (Sackett and Sparke 1990, Sackett 1991)
and of HI warps in spiral disks
(e.g., Hofner \& Sparke 1991, Casertano 1991).
In particular, it has been suggested that warps are
bending modes (Sparke \& Casertano 1988), in which a disk precesses into a
warped configuration because
it is initially tilted with respect to the symmetry
plane of an oblate halo (Toomre 1983, Dekel \& Shlosman 1983; for
recent discussions see Casertano 1991, Hofner \& Sparke 1991, Binney 1992).
So far, this idea is in reasonable agreement with
warp observations (Briggs 1990, Bosma 1991),
predicting both a straight line of nodes in
the inner parts of disks and that concentrated halos inhibit
warping. Although complex issues, such as the halo
response to the changing potential of the disk, require further study
(Casertano 1991, Binney 1992),
misalignment between the symmetry axes of flattened
halos and disks remains a promising scenario for explaining
warps. We also note that
oblate halos with misaligned, warped disks
appear to be a common outcome of
N-body simulations of galaxy formation in cold dark matter models
(Dubinski \& Carlberg 1991, Katz \& Gunn 1991) and of galaxy
mergers (Hernquist 1989), which might have
relevance here if MACHOs form sufficiently early in the history of the galaxy.

Here we study microlensing for an axisymmetric
flattened halo model (Binney \& Tremaine 1987),
taking into account the anisotropy of
the velocity dispersion consistent with the halo density
distribution (Evans 1993). Since
the Milky Way is known to be warped beyond the solar circle
(Henderson, Jackson, \& Kerr 1982),
we consider an oblate halo whose plane of symmetry does not necessarily
coincide with the galaxy disk plane. Bearing in mind the complexity of
the Milky Way
warp (e.g., in the south, the HI disk curves back on itself), we consider
a moderate range of misalignment angles between disk and halo. In contrast with
the case of a flattened halo coplanar with the disk (Sackett \& Gould 1994),
allowance for disk-halo misalignment significantly widens the spread in
the LMC microlensing optical depth $\tau(LMC)$ and makes the
$\tau_{SMC}/\tau_{LMC}$ ratio test for halo flattening, while still
potentially useful, significantly less robust
(more ambiguous). These effects are easily understood on geometric grounds.

\vskip 2pc
\head{2. The Halo Model and Rotation Curve Constraints}
\taghead{2.}
\vskip 1pc

We consider an axisymmetric flattened halo derived from the logarithmic
potential (Binney 1981, Binney \& Tremaine 1987)
$$\Phi = -{ v_0^2 \over 2} \ln \Big(R_c^2+R^2+z^2q^{-2} \Big) ~,
\eqno(phi)$$
where ($R,z,\phi$) are cylindrical coordinates,
$R_c$ is the core radius, $q$ is the axis ratio of the
spheroidal equipotentials, and $v_0$ is the rotation velocity at
infinite radius in the equatorial plane ($z=0$) of the halo.
The corresponding density profile is
given by
$$\rho(R,z)= {v_0^2 \over {4\pi G q^2}}
{{(2q^2+1)R_c^2+R^2+(2-q^{-2})z^2} \over {(R_c^2+R^2+z^2q^{-2})^2}} ~.
\eqno(den)$$
The asymptotic axis
ratio of the equidensity contours is $1:1:q^{-3}$. Note that in the
limit $R_c \rightarrow 0$ and $q \rightarrow 1$, \(den) reduces
to a singular isothermal sphere.
The unique even part of the distribution function $f(E,L_z^2)$ can be
recovered from $\rho (R,\Phi)$ by a double Laplace
inversion (Lynden-Bell 1962) or by contour integration techniques (Hunter
\& Quian 1993) to give (Evans 1993):
$$f(E,L_z^2)= (A L_z^2+B) \exp (4E/v_0^2) + C \exp (2E/v_0^2) ~,
\eqno(f)$$
where $E= \Phi - v^2/2$, $L_z = R v_\phi$, and the parameters
$A\equiv (2/\pi)^{5/2} (1-q^2)/(Gq^2v_0^3)$, $B\equiv
(2/\pi^5)^{1/2} (R_c^2 / Gq^2v_0)$ and $C\equiv
(2q^2-1)/(4\pi^{5/2}Gq^2v_0)$. Adding arbitrary terms odd in the $z$-component
of angular momentum $L_z$ to
this distribution function does not change the density, but determines
the total angular momentum of the halo. The odd part of the
distribution function can be determined uniquely given a rotation law
$<v_\phi>(R)$ (Lynden-Bell 1962, Evans 1993). In this letter we consider
the case of no net
streaming, so the distribution function is given by its even part
\(f); this is positive definite as long as $0.707\leq q \leq 1.08$.
We focus on the case of oblate halos, with $q \leq 1$.

To apply this halo model to the Milky Way, we must
consider the restrictions on the parameters $R_c$
and $v_0$ imposed by the observed rotation curve of the
Galaxy. The contribution of the halo to
the rotation velocity in the disk plane is found from \(phi),
$$v= v_0 \Big({r^2 \over {a^2 + r^2}}\Big)^{1/2} ~,\eqno(rc)$$
where $r$ is the radial distance in the plane of the disk,
$a^2=R_c^2 / (q^{-2} \sin^2 \theta + \cos^2 \theta)$, and $\theta$
is the tilt angle between the symmetry plane of the halo and the
disk. To find the rotation curve, we add the halo contribution \(rc) in
quadrature with those from the
visible components of the Galaxy, i.e., the disk, the spheroid, and
a central component. For the visible components, we use the model of
Bahcall,
Schmidt \& Soneira (Bahcall, Schmidt, \& Soneira 1982, 1983, Bahcall
and Soneira 1980).
This corresponds approximately to the `light disk'
model used in SG; our conclusions would not change
qualitatively if we considered a range of models for the visible components.
The rotation curve of the Milky Way is known to be
approximately flat between 3 and 16 kpc (for a review, see
Fich and Tremaine 1991); we impose the constraint that the circular velocity
$v(r)$ in the disk lie between
200 and 240 km/sec over the interval $r= 3 - 16$ kpc (this is similar but
not identical to the constraint imposed by SG).
This leads to an allowed region in the $(a,v_0)$ plane contained within
the bounds
$0 < a < 7.7$ kpc and $140$ km/sec $< v_0 < 214$ km/sec.
The shape of the constraint region is such that lower values of $a$ are
paired with lower values of $v_0$, and large $a$ with higher $v_0$.

If the halo is tilted with respect to the disk, the second relevant
parameter describing the halo orientation, in addition to the tilt
angle $\theta$, is the angle $\psi$ between
the ``unobservable'' line of nodes (the intersection of
the halo symmetry plane with the disk) and the sun-galactic
center line (SGCL). We set this angle to be the observed angle between
the SGCL and the ``observable'' line of nodes (the intersection of the
plane of the outer disk with that of the inner disk), which is about
10 degrees due north (Henderson, Jackson, \& Kerr 1982). We have
also investigated other choices of $\psi$; the choice above happens
to have roughly the largest impact on
the microlensing rates, so results for other $\psi$ will be bracketed
by those shown below.

We must choose a reasonable range of values for the tilt angle $\theta$.
The warp angle, defined as the angle
between the plane of the outer disk and that of the inner disk, is
about 18 degrees for the Milky Way (Henderson, Jackson, \& Kerr 1982).
In the ``modified tilt mode" model (Sparke and Casertano 1988),
the relation between the
warp angle and the unobservable tilt angle $\theta$
depends on the parameters of the halo and the disk.
Roughly speaking, for small $R_c$ (in units of the disk scale length),
the warp angle increases with radius (``Type I'' mode),
whereas in the
opposite case the warp angle decreases with radius (``Type II''
mode). The parameters of the Milky Way naively appear to favor the
Type I mode, but the complexity of the Galactic warp and the simplicity
of the model mitigate against giving this much weight. (The constraint
that the mode be discrete in principle constrains the core radius and
halo ellipticity, but this does not constrain the halo parameter space
more strongly than the rotation curve above.)
For a Type I mode, $\theta$ can be smaller or larger than the
warp angle (depending on the halo mass compared to the disk mass, and
on the flattening). For a Type II mode, $\theta$ is always greater than
the warp angle. Therefore, we consider values of $\theta$ between -30 and
+30 degrees to cover a plausible range.
Positive values of $\theta$ denote tilting the halo symmetry
plane ``towards the LMC'', whereas negative values correspond to
tilting ``away from the LMC''.  This geometric picture accounts qualitatively
for the results below.

\vskip 2pc
\head{3. Microlensing Results}
\taghead{3.}
\vskip 1pc

We now use the constrained halo models above to study microlensing.
The optical
depth is given by the number of MACHOs inside the ``microlensing
tube'' (Griest 1991):
$$\tau = \int_0^{x'_hL} {\rho(R(x),z(x)) \over m} \pi u_T^2 R_E^2(x)
dx ~, \eqno(tau)$$
where $m$ is the MACHO mass, $R_E=2(Gmx(L-x)/c^2L)^{1/2}$ is the
Einstein ring radius, $L$ is the distance to the lensed star, $u_T$ is the
threshold impact parameter in units of the Einstein radius, $x$ denotes
the distance along the line of sight, and $x'_hL$ is the lesser of
the extent of the halo and $L$. Studies using high velocity stars to infer
the local escape speed (Fich \& Tremaine 1991)
suggest that the truncation radius of the halo is in excess of
35 kpc; therefore, since 90\% of the
microlensing occurs between 5 and 30 kpc, extending the halo out to the
LMC or SMC will introduce negligible error. We thus take $x'_h$ to be
1. (It would be useful to check this by using the physically truncated
lowered Evans models of Kuijken and Dubinski 1993).

In Figure 1, we show the LMC optical depth
$\tau_{LMC}$ (assuming $l=280.5^o$, $b=-32.9^o$,
and $L=50$ kpc) as a function of core radius, for different values of
the halo-disk tilt angle $\theta$ for an E6 model
(with asymptotic axis ratio equal to 0.4, and $q=0.737$). For comparison, we
show the spherical halo (E0) values (solid curves). For each
$\theta$, we show the maximum and minimum values of $\tau_{LMC}$
allowed by the rotation curve constraints.
Note that the LMC optical depth is insensitive to halo flattening without
tilt (the E0 and E6 $\theta=0$ curves are nearly identical), and for
$\theta=0$
the allowed range in optical depth is $\tau(LMC) \simeq 3 - 6 \times 10^{-7}$,
both in good agreement with SG. On the other hand,
the LMC optical depth is quite sensitive to the tilt angle:
$\tau(LMC)$ increases
with $\theta$, reflecting the increasing mass between us and the LMC.
As a result, for $-30^o < \theta < 30^o$, the spread in
$\tau_{LMC}$ is increased to an order of magnitude, $\tau(LMC) \simeq
1.2 - 10 \times 10^{-7}$.

As pointed out by SG, the ratio
$\tau_{SMC}/\tau_{LMC}$ is independent of the rotation curve
constraints (because $\rho \propto v_0^2$). In Figure 2, we show the ratio
$\tau_{SMC}/\tau_{LMC}$ as a function of $R_c$ for each
$\theta$ (we assume $l=302.8^o$, $b=-44.3^o$, and $L=63 kpc$ for the
SMC). For the spherical E0 model, we find the ratio is larger than 1.45,
while for the untilted E6 model it is below 1.0, in agreement with SG.
This large difference was
the basis of SG's proposal to use this ratio as a test of halo ellipticity.
However, for $\theta=30^o$, the E6 model ratio is about
1.22-1.24, significantly closer to the E0 value of 1.46. SG estimate that
the fractional precision in this ratio measurement would be at best
10 \% due to statistical fluctuations, and perhaps a factor of two
larger, depending on the extent to which MACHOs in the LMC contaminate and
can be removed from the sample. Consequently, there appears to be
some degeneracy in the ratio test: a measurement of $\tau_{SMC}/\tau_{LMC}$
in the range $1.2 - 1.4$ would not reliably determine the
shape of the Galaxy halo. Nevertheless, it would still help
narrow down the range of possibilities in the ellipticity-orientation
parameter space. On the other hand, a measurement of the ratio which
came out below unity would definitely point to halo flattening, although
it would provide little information on the tilt, which is unfortunate
given the insights this might provide on the dynamics of warps.

Recently, Gould, Miralda-Escud\'e, and Bahcall (1993) have discussed
tests to distinguish microlensing by disk and halo populations. For
MACHOs in a disk, they find $\tau_{SMC}/\tau_{LMC} = 0.6$, substantially
below their values of 1.47 for a spherical halo and 0.96 for a
flattened coplanar E6 halo. We find that this ratio can be as low
as 0.84 for an E6 halo with $\theta = -15^o$, so this particular test can still
potentially discriminate between halo and disk populations, although
with less confidence.

Another quantity of interest is the microlensing event rate $\Gamma$,
the rate at which MACHOs enter the microlensing
tube (Griest 1991):
$$\Gamma= \int {f({\bf v},R(x),z(x)) \over m} v_r^2 \cos \phi u_T R_E(x) dv_r
dv_x d\phi dx d\alpha ~. \eqno(gamma)$$
Here $v_r$ is the transverse velocity of the MACHOs in the frame of
the microlensing tube, $\phi$ is the angle between $v_r$ and the
normal to the surface of the tube, and $\alpha$ is the polar angle in
the plane normal to the line of sight (see Griest 1991; we do not
include the motion of the Earth or of the LMC). The microlensing rate
for the LMC looks qualitatively similar to the optical depth results of
Fig. 1: there is an order of magnitude spread in the values of $\Gamma_{LMC}$,
over the range $0.25 - 2.9 \times 10^{-6} (m/M_\odot)^{-1/2}u_T$ events/yr.
In Figure 3 we show the ratio $\Gamma_{SMC}/\Gamma_{LMC}$ for
different values of $\theta$ as a function of $R_c$ for E6 and E0
halos, taking into account the rotation curve constraints. The
dependence of the rate ratio on $R_c$ differs from that for the optical
depth, because the velocity dispersion is tied to the other halo parameters
through \(f) and the rotation curve; the velocity anisotropy plays a
relatively minor role.
But the dependence of the rate ratio on halo-disk tilt angle
is qualitatively similar to that for the optical depth. Note that, in
the absence of independent information on the halo core radius,
the degeneracy between the spherical and flattened tilted models is
more severe for the rate ratio than for the optical depth ratio.

Finally, we calculate the relative probability $P(m,t_e)$
of models characterized by
mass $m$ giving rise to an event duration of $t_e$.
Assuming that all the MACHOs have the same mass, this is
given by (Griest 1991):
$$P(m,t_e) = {\Delta t_e\over \Gamma} \int {f\over m} (v_0^2 \beta)^2 y^{3/2}
(1-y)^{-1/2} dv_x dy dx d\alpha,\eqno(mass)$$
where $v_0^2 \beta \equiv (2R_E u_T/t_e)^2$, and $\beta y \equiv
v_r^2/v_0^2$. The constant $\Delta t_e$ is chosen so that the maximum of
$P(m,t_e)$ is 1. For comparison with the results of Griest (1991), we
show this quantity in Fig. 4 for $t_e = 0.3$ yr, assuming $u_T=1$ and
$R_c = 4$ kpc.
The mass scales as $(t_e/0.3)^2/u_T^2$. The results show that the
inferred mass is relatively insensitive to halo flattening or tilt, but
that the allowed spread in halo velocity dispersion at fixed $R_c$ gives
rise to an additional factor of $\sim 1.6$ uncertainty in the mass.

\vskip 2pc
\head{4. Conclusion}
\vskip 1pc

Motivated in part by models of HI disk warps and by N-body simulations of
galaxy formation, we have studied microlensing in an oblate halo that is
not coplanar with the disk. Imposing constraints from the observed
Galaxy rotation curve, we find that the lensing optical depth for the
LMC is sensitive to the tilt angle between disk and halo, and can be
twice as large for moderate tilt angle as for coplanarity. The resulting
spread in LMC optical depth is roughly an order of magnitude, $\tau(LMC)
\simeq 1- 10 \times 10^{-7}$ for models consistent with the flatness of
the rotation curve. The tilt angle also affects the ratio of the
optical depth to the SMC and LMC, rendering it a less robust test
of halo flattening. While a
small ratio, $0.8 < \tau_{SMC}/\tau_{LMC} < 1$
would unambiguously indicate flattening, a larger ratio does not cleanly
indicate sphericity: there is a near-degeneracy for spherical and flattened
tilted halos. On the other hand, a
very large optical depth for the LMC, $\tau > 7\times 10^{-7}$,
would help break this degeneracy and would point to a flattened halo with
tilt.

We can envision a number of directions which could be taken to improve
upon or extend the results discussed here. These include the study of
triaxial halos, self-consistent modelling of the disk-halo system (the
halo model discussed here is an equilibrium solution if the gravitational
field of the disk is neglected), and the exploration of other
observational tests of disk-halo tilt in the Milky Way.

We thank E. Gates and M. Turner for useful conversations. This research
was supported in part by the DOE and by NASA grant NAGW-2381 at Fermilab.

\vskip 2pc
\head{References}
\vskip 1pc

\noindent Alcock, C., et al.,\nat 365,621,1993..\par

\noindent Ashman, K., \pasp 104,1109,1992..\par

\noindent Aubourg, E., et al., \nat 365,623,1993..\par

\noindent Bahcall, J. N. \& Soneira, R. M., \apjs 44,73,1980..\par

\noindent Bahcall, J. N., Schmidt, M., \& Soneira, R. M.,
\apj 258,L23,1982..\par

\noindent Bahcall, J. N., Schmidt, M., \& Soneira, R. M.,
\apj 265,730,1983..\par

\noindent Binney, J. J., \mnras 196,455,1981..\par

\noindent Binney, J. J., \araa 30,51,1992..\par

\noindent Binney J. J., \& Tremaine, S., {\it Galactic Dynamics},
Princeton University Press (1987).\par

\noindent Bosma, A., in {\it Warped
Disks and Inclined Rings around Galaxies}, Cambridge University Press,
181 (1991).\par

\noindent Briggs, F. H., \apj 352,15,1990..\par

\noindent Casertano, S., in {\it Warped Disks and Inclined Rings
around Galaxies}, Cambridge University Press, 237 (1991).\par

\noindent Dekel A., \& Shlosman, I., in {\it IAU Symp. 100, Internal Kinematics
\& Dynamics
of Galaxies}, ed. E.Athanassoula, Dordrecht, Reidel, 177 (1983).\par

\noindent Dubinski, J. \& Carlberg, R. G., \apj 378,496,1991..\par

\noindent Evans, N. W., \mnras 260,191,1993..\par

\noindent Fich, M. \& Tremaine, S., \araa 29,409,1991..\par

\noindent Gates, E., \& Turner, M. S., Fermilab preprint
Fermilab-Pub-93/357-A.\par

\noindent Gould, A., Miralda-Escud\'e, J., \& Bahcall, J. N.,
Institute for Advanced Study preprint IASSNS-AST 93/63 (1993).\par

\noindent Griest, K., \apj 366,412,1991..\par

\noindent Hofner, P. \& Sparke, L. S., in {\it Warped Disks and
Inclined Rings around Galaxies}, Cambridge University Press, 225 (1991).\par

\noindent Henderson, A. P., Jackson, P. D., \& Kerr, F. J., \apj
116,122,1982.. \par

\noindent Hernquist, L., \nat 340,687,1989..\par

\noindent Hunter, C. \& Quian, E., \mnras 262,401,1993..\par

\noindent Katz, N. \& Gunn, J., \apj 377,365,1991..\par

\noindent Kuijken, K. \& Dubinski, J., preprint, (1993). \par

\noindent Lynden-Bell, D., \mnras 123,447,1962..\par

\noindent Paczy\'nski, B., \apj 304,1,1986.. \par

\noindent Sackett, P. D., in {\it Warped Disks and
Inclined Rings around Galaxies}, Cambridge University Press, 73 (1991).\par

\noindent Sackett, P. D.  \& Gould, A. \apj 419,in press,1994..\par

\noindent Sackett, P. D. \& Sparke, L. S. \apj 361,408,1990..\par

\noindent Sparke, L. S., \& Casertano, S., \mnras 234,873,1988..\par

\noindent Toomre, A. in {\it IAU Symp. 100, Internal Kinematics \& Dynamics
of Galaxies}, ed. E.Athanassoula, Dordrecht, Reidel, 177 (1983).\par

\noindent Udalski, A., et al., {\it Acta Astron.} {\bf 43}, 289, (1993).\par

\vskip 2pc
\head{Figure Captions}
\vskip 1pc

\noindent {\bf Figure 1.} LMC optical depth as a function of halo core radius
$R_c$
for spherical E0 and flattened E6 halos, assuming $u_T =1$
(for other values, $\tau \propto u_T^2$). For each tilt angle $\theta$, we
show the maximum and minimum values of the optical depth consistent with the
rotation curve constraints. \vskip 1pc

\noindent {\bf Figure 2.} Ratio of optical depth toward the Small and Large
Magellanic
clouds, $\tau_{SMC}/\tau_{LMC}$, as a function of core
radius for E6 halos (with different tilt angles) and the spherical E0 halo.
\vskip 1pc

\noindent {\bf Figure 3.} Ratio of microlensing rates for the Small and Large
clouds,
$\Gamma_{SMC}/\Gamma_{LMC}$, as a function of core radius,
for E6 halos (with different tilt angles) and the E0 halo.

\vskip 1pc

\noindent {\bf Figure 4.} Relative probability of models characterized by
MACHO mass $m$
giving rise to an event duration $t_e=0.3$ yr, assuming $u_T=1$ and
halo core radius $R_c = 4$ kpc. The curves denote the same
models as shown in Fig. 1, and the two curves for each model correspond
to the maximum and minimum halo velocities for this core radius.

\end